\documentclass[12pt]{article}
\usepackage{amstex,amssymb,epsfig,cite}

\numberwithin{equation}{section}

\begin{document}
\begin{flushright}
SU-4240-701 \\
TIFR/TH/99-32 \\
hep-th/9910129\\
\end{flushright}

\begin{center}
{\large{\bf INSTANTONS AND CHIRAL ANOMALY IN FUZZY PHYSICS}}

\bigskip 
A. P. Balachandran$^a$ and S. Vaidya $^b$
\footnote{sachin@@theory.tifr.res.in} \\ 
$^a${\it Physics Department, Syracuse University, \\
Syracuse,N.Y.,13244-1130, U.S.A.}

$^b${\it Tata Institute of Fundamental Research, \\
Colaba, Mumbai, 400 005, India.}

\end{center}

\begin{abstract}
In continuum physics, there are important topological aspects like
instantons, $\theta$-terms and the axial anomaly. Conventional lattice
discretizations often have difficulties in treating one or the other
of these aspects. In this paper, we develop discrete quantum field
theories on fuzzy manifolds using noncommutative geometry. Basing
ourselves on previous treatments of instantons and chiral fermions
(without fermion doubling) on fuzzy spaces and especially fuzzy
spheres, we present discrete representations of $\theta$-terms and
topological susceptibility for gauge theories and derive axial anomaly
on the fuzzy sphere. Our gauge field action for four dimensions is
bounded by a constant times the modulus of the instanton number as in
the continuum.
\end{abstract}

\section{Introduction}
Conventional discretizations of quantum fields on a manifold ${\cal
M}$ replace the latter by a lattice of points. An alternative
discretization which leads to fuzzy physics treats ${\cal M}$ as a
phase space and quantizes it. ${\cal M}$ is thereby altered to a
'fuzzy' manifold ${\cal M}_F$
\cite{madore,gropre,grklpr1,grklpr2,grklpr3,watamura1,watamura2,frgrre}.
Earliest investigations of quantization like those of Planck and Bose
show that quantization introduces a short distance cut-off: the number
of states in a phase space volume $V$ is reduced from infinity to
$V/\tilde{h}^3$ if $\tilde{h}$ plays the role of Planck's constant. If
${\cal M}$ is compact, the total number of states is also finite and
we end up with a matrix model for ${\cal M}$. Continuum physics in
this approach has to do with the `classical' limit $\tilde{h}
\rightarrow 0$. Functions on ${\cal M}$ commute, but they become
noncommutative on quantization. For that reason, the fuzzy path lands
us in noncommutative manifolds and their geometries
\cite{madore,connes,landi,coquereaux,vargra,mssv,varilly}. As there
are also reasonably orderly methods to formulate quantum field
theories (QFT's) on ${\cal M}_F$, fuzzification promises to be a truly
original development in discrete physics.

The earliest contributions to topological features of fuzzy physics
came from Grosse, Klim\v{c}\'{\i}k and Pre\v{s}najder
\cite{grklpr1}. They dealt with monopoles and chiral anomaly for the
fuzzy two-sphere $S_F^2$ and took particular advantage of
supersymmetry. Later we \cite{bbiv} further elaborated on their
monopole work and also developed descriptions of fuzzy $\sigma$-models
and their solitons using cyclic cohomology \cite{connes,coquereaux} in
an important manner. An attractive feature of this cohomological
approach is its ability to write analogues of continuum winding number
formulae and derive a fuzzy Belavin-Polyakov bound \cite{bbiv}.

The work of \cite{bbiv} is extended in this paper to gauge theories
and chiral anomalies. Instantons, the $\theta$-term and ``topological
susceptibility'' are in particular formulated on $S_F^2$. They are not
like the existing proposals in discrete physics (cf. the work of
L\"{u}scher reported in ref. \cite{MandM}), have sound and rugged
interpretations and promise to resolve problems of much age. As for
chiral fermions and anomaly on $S_F^2$, it is known that fuzzy physics
requires no ``fermion doubling'' for Watamuras' Dirac operator
\cite{watamura1,watamura2}.  Unfortunately it has zero modes, also
there is another Dirac operator
\cite{gropre,grklpr1,grklpr2,grklpr3,watamura1,watamura2} which gives
a much better approximation to the continuum spectrum. But it does not
have a chirality operator because of its highest frequency mode. This
mode recedes to infinity and is totally unimportant in the continuum
limit. In \cite{ongoing}, it is projected out and thereafter a
chirality operator anticommuting with this Dirac operator is found
{\it without} fermion doubling. This paper adapts this operator to
gauge theories and instanton physics and also derives the axial
anomaly. An alternative analysis of the latter can be found in Grosse
et al \cite{grklpr1}. We will briefly review their work and compare it
with ours in the final section. An interesting new treatment of chiral
fermions and anomaly is also discussed in \cite{luscher}.

Not all manifolds ${\cal M}$ can be phase spaces. ${\cal M}$ has to be
symplectic and hence even dimensional. It must be quantizable to turn
into ${\cal M}_F$, and ideally ${\cal M}$ must be compact and ${\cal
M}_F$ admit a Laplacian and Dirac operator with decent symmetry
properties.  Manifolds with these nice features are quantizable
coadjoint orbits of compact Lie groups. For simple and semi-simple Lie
groups, they are also adjoint orbits.  Examples are ${\mathbb
C}{\mathbb P}^N$ and $S^2 \times S^2$ ($S^2$ being ${\mathbb
C}{\mathbb P}^1$). [$T^2$ is not in this category.] We focus on $S^2$
and $S^2 \times S^2$ in this paper as the basics of $S_F^2$ are under
reasonable control. ${\mathbb C}{\mathbb P}^2$ has been treated by
\cite{grostr} while our own approach to ${\mathbb C}{\mathbb P}^N$ and
other orbits is yet to appear in print. Using $S^2$ and $S^2 \times
S^2$ as examples, we present our considerations in such a manner that
they can be easily be adapted to general adjoint orbits once their
fuzzy basics are assumed.

\section{The Fuzzy Sphere}
A sphere $S^2$ is a submanifold of ${\mathbb R}^3$:
\begin{equation} 
S^2 = \langle \vec{n} \in {\mathbb R}^3: \sum_{i=1}^3 n_i^2=1 \rangle. 
\end{equation} 
If $\hat{n}_i$ are the coordinate functions on $S^2$,
$\hat{n}_i(\vec{n}) = n_i$, then $\hat{n}_i$ commute and the algebra
${\cal A}$ of smooth functions they generate is commutative.

In contrast, the operators $x_i$ describing $S_F^2$ are
noncommutative:
\begin{equation} 
[x_i, x_j] = \frac{i \epsilon_{ijk} x_k}{[l(l+1)]^{1/2}}, \quad
\sum_{i=1}^3 x_i^2 = {\bf 1}, \quad l \in \left\{ \frac{1}{2}, 1, 
\frac{3}{2}, \ldots \right\}. 
\end{equation} 
The $x_i$ approach $\hat{n}_i$ as $l \rightarrow \infty$. If $L_i =
[l(l+1)]^{1/2}x_i$, then $[L_i, L_j] = i \epsilon_{ijk}L_k$ and $\sum
L_i^2 = l(l+1)$ so that $L_i$ give the irreducible representation
(IRR) of $SU(2)$ Lie algebra for angular momentum $l$. $L_i$ or $x_i$
generate the algebra $A=M_{2l+1}$ of $(2l+1) \times (2l+1)$ matrices.

Scalar wave functions on $S^2$ come from elements of ${\cal A}$. In a
similar way, elements of $A$ assume the role of scalar wave functions
on $S_F^2$. A scalar product on $A$ is $\langle \xi, \eta \rangle = Tr
{\xi}^{\dagger} \eta$. $A$ acts on this Hilbert space by left- and
right- multiplications giving rise to the left and right- regular
representations $A^{L,R}$ of $A$. For each $a \in A$, we thus have
operators $a^{L, R} \in A^{L,R}$ acting on $\xi \in A$ according to
$a^L \xi = a \xi, a^R \xi = \xi a$. [Note that $a^L b^L = (ab)^L, a^R
b^R = (ba)^R$.] We assume by convention that elements of $A^L$ are to
be identified with functions on $S^2$. 

Of particular interest are the angular momentum operators. There are
two kinds of angular momenta $L_i^{L,R}$ for $S_F^2$, while the
orbital angular momentum operator, which should annihilate ${\bf 1}$
is ${\cal L}_i = L_i^L - L_i^R$. $\vec{\cal L}$ plays the role of the
continuum $-i(\vec{r} \times \vec{\nabla})$. The ``position''
operators are not proportional to ${\cal L}_i$, but are instead
$L_i^L/[l(l+1)]^{1/2}$.

The elements of $A$ have a dual role, one as members of a Hilbert
space and the second as operators on this space. We shall hereafter
most often denote these Hilbert space vectors and their duals in a
ket-bra notation to minimize confusion. Thus $|\eta \rangle$ is the
Hilbert space vector for $\eta \in A$, $\langle \xi|\eta \rangle = Tr
{\xi}^{\dagger} \eta, a^L |\xi \rangle = |a \xi \rangle$ and $ a^R|\xi
\rangle = |\xi a \rangle$.

There are two Dirac operators ${\cal D}_\alpha$ on $S^2$ that are of
particular importance to us:
\begin{eqnarray} 
{\cal D}_1 &=& \vec{\sigma}. [-i(\vec{r} \times \vec{\nabla})] + {\bf
1}, \\
{\cal D}_2 &=& -\epsilon_{ijk}\sigma_i \hat{n}_j {\cal J}_k,
\end{eqnarray} 	 
where 
\begin{equation} 
{\cal J}_k = [-i(\vec{r} \times \vec{\nabla})]_k + \sigma_k/2 =
\text{Total angular momentum operators.}
\end{equation} 
There is a common chirality operator $\Gamma$ anticommuting with both:
\begin{eqnarray}	             
\Gamma = \vec{\sigma}.\hat{n} &=& \Gamma^{\dagger}, \quad \Gamma^2 =
{\bf 1}, \\
\Gamma {\cal D}_\alpha + {\cal D}_\alpha \Gamma &=&0.
\end{eqnarray} 
${\cal D}_\alpha$ and $\Gamma$ act on spinors $\psi = (\psi_1,
\psi_2)$ where $\psi_i \in A$. Also these Dirac operators {\it in the
continuum} are unitarily equivalent,
\begin{equation}
{\cal D}_2 = \exp{(i \pi \Gamma/4)} {\cal D}_1 \exp{(-i \pi \Gamma/4)}
\end{equation}
and have the spectrum 
\begin{equation}
\text{Spectrum of}\,\,{\cal D}_\alpha =  \{ \pm (j+1/2): j \in \{1/2,
3/2, \ldots \} \}, 
\end{equation}
where $j$ is total angular momentum (spectrum of $\vec{\cal J}^2 =
\{j(j+1)\}$ ). There is a circle of possibilities $\{ e^{(i \theta
\Gamma/2)} {\cal D}_1 e^{(-i \theta \Gamma/2)} \}$, in which the
operators ${\cal D}_\alpha$ are just two points.

The discrete versions of ${\cal D}_\alpha$ are
\begin{eqnarray} 
D_1 &=& \vec{\sigma}. \vec{\cal L} + {\bf 1}, \\ 
D_2 &=& -\epsilon_{ijk}\sigma_i x^L_j J_k = \epsilon_{ijk}\sigma_i 
        x_j^L L_k^R,
\end{eqnarray} 
where
\begin{equation} 
J_k = {\cal L}_k + \sigma_k/2 =\text{Total angular momentum
operators.} \label{tamo}
\end{equation} 
The $D_\alpha$ are no longer unitarily equivalent, their
spectra being 
\begin{eqnarray}
\text{Spectrum of}\,\,D_1 &=&  \left\{ \pm (j+\frac{1}{2}): j \in
                              \{ \frac{1}{2}, \frac{3}{2}, \ldots
                              2l-\frac{1}{2} \} \right\} \nonumber \\  
                        & & \cup \left\{ (j+\frac{1}{2}):
                              j=2l+\frac{1}{2} \right\},\label{specd1}\\
\text{Spectrum of}\,\,D_2 &=&  \left\{ \pm(j+1/2)\left[1 +
                              \frac{1-(j+1/2)^2}{4l(l+1)}\right]^{1/2}:
                              \nonumber \right.\\
& & \!\!j \left. \in \{ \frac{1}{2},\frac{3}{2}, \ldots 2l-\frac{1}{2} \}
                              \right\} \cup \left\{0: j= 2l +
                              \frac{1}{2} \right\}\!\!.\label{specd2}
\end{eqnarray}
$j$ once more is total angular momentum, the spectrum of $\vec{J}^2$
being $\{j(j+1)\}$.

The first operator has been used extensively by Grosse et al
\cite{grklpr1,grklpr2,grklpr3} while the second was first introduced
by the Watamuras \cite{watamura1,watamura2}.

The Hilbert space $A$ has naturally to be enhanced to $A^2 = A \otimes
{\mathbb C}^2$ once the Dirac operator comes into the picture. It is
spanned by vectors $\{|\xi \rangle : \xi=(\xi_1, \xi_2), \xi_i \in A
\}$ with the scalar product $\langle \xi | \eta \rangle = Tr
\xi^{\dagger}_i \eta_i$. $A^{L,R}$ act in an obvious manner on this
space, $a^L |\xi \rangle = |a \xi \rangle, a^R|\xi \rangle = |\xi a
\rangle$, while $\sigma_i |\xi \rangle$ is just $|\sigma_i \xi
\rangle$. Here $(a \xi)_r =a \xi_r, (\xi a)_r = \xi_r a$ and
$(\sigma_i \xi)_r = (\sigma_i)_{rs} \xi_s$.

It is easy to derive (\ref{specd1}) by writing 
\begin{eqnarray}
D_1&=& \left( \vec{\cal L} + \frac{\vec{\sigma}}{2}\right)^2 -
\vec{\cal L}^2 - \left(\frac{\vec{\sigma}}{2}\right)^2 + {\bf 1}, \\
\left(\frac{\vec{\sigma}}{2}\right)^2 &=& \frac{3}{4} {\bf 1}.
\end{eqnarray}
Then for $\vec{\cal L}^2 = k(k+1), k \in \{0, 1, \ldots 2l \}$, if
$j=k+1/2$ we get $+(j+1/2)$ as eigenvalue, while if $j=k-1/2$ we get
$-(j+1/2)$. The absence of $-(2l+1/2)$ in (\ref{specd1}) is just
because $k$ cuts off at $2l$.

It is remarkable that (\ref{specd1}) coincides {\it exactly} with
those of ${\cal D}_\alpha$ upto $j=(2l-1/2)$. So $D_1$ is an excellent
approximation to ${\cal D}_\alpha$.

But $D_1$ as it stands admits no chirality operator unless its
eigenspace with top eigenvalue is treated as a chiral {\it singlet},
or better still is {\it projected out}. Then it {\it does} admit a
chirality operator as shown in detail in \cite{ongoing}. We can
summarize the results of \cite{ongoing} as follows. Define the
operators
\begin{eqnarray}
\epsilon_\alpha &=& \frac{D_\alpha}{|D_\alpha|}, \quad |D_\alpha| =
                                              \text{Positive square 
                                              root of}\,\, D_\alpha^2,
                                              \nonumber \\ 
& &\quad\quad\quad \quad\text{on the subspace $V$ with}\;j \leq 2l-1/2,
                                              \nonumber \\ 
 &=& 0 \quad \text{on the subspace $W$ with}\;j=2l+1/2.
\end{eqnarray}
Then $\epsilon_1, \epsilon_2$ and $i \epsilon_1\epsilon_2$
anticommute, are hermitian and square to the projection operator $P$
where 
\begin{eqnarray} 
P \xi &=& \xi, \quad \xi \in V, \nonumber \\
      &=& 0,  \quad \xi \in W.
\end{eqnarray}
Furthermore they all commute with $|D_\alpha|$ so that $i
\epsilon_1\epsilon_2$ is a chirality operator for $D_\alpha$.

As for $D_2$, its spectrum (\ref{specd2}) has been calculated in
\cite{watamura1,watamura2}. It has the following chirality too,
anticommuting with $D_2$, commuting with $\vec{J}$ and squaring to
${\bf 1}$ in the {\it entire} Hilbert space $V \oplus W = A^2$:
\begin{eqnarray} 
\gamma_2 &=& \gamma_2^{\dagger} = -\frac{\vec{\sigma}.{\vec{L}}^R
-1/2}{l+1/2}, \\
\gamma_2^2 &=& {\bf 1}.
\end{eqnarray} 
As shown in \cite{ongoing}, on $V$, it is a linear combination of
$\epsilon_1$ and $i\epsilon_1\epsilon_2$ for each fixed $\vec{J}^2
\equiv J^2$:
\begin{eqnarray} 
P\gamma_2 P &=& e^{(i \theta (J^2) \epsilon_2)/2}
(i\epsilon_1\epsilon_2) e^{(-i \theta (J^2) \epsilon_2)/2}, \nonumber \\
   &=& \cos \theta (J^2) (i\epsilon_1\epsilon_2) + \sin \theta (J^2)
\epsilon_1. 
\end{eqnarray}

The angle $\theta$ is rotationally invariant and vanishes as $l
\rightarrow \infty$. In that limit, $\gamma_2$ and $P\gamma_2 P$
approach the continuum $\Gamma$ as they should. 

The operator $\gamma_2$ has the important property that it commutes
with ${x_i}^L$. That is not the case with $i\epsilon_1\epsilon_2$. It
is thus useful to replace $D_1$ by
\begin{equation} 
e^{(i \theta (J^2) \epsilon_2)/2} D_1 e^{(-i \theta
(J^2) \epsilon_2)/2} \nonumber.
\end{equation} 
Noticing also that $P$ is a function of only $\vec{J}^2$ and that
$\vec{J}^2$ commutes with $\gamma_2$ and $D_\alpha$, we construct our
basic operators 
\begin{eqnarray} 
D &=& e^{(i \theta (J^2) \epsilon_2)/2} (P D_1 P) e^{(-i \theta (J^2)
\epsilon_2)/2}, \\
F &=& e^{(i \theta (J^2) \epsilon_2)/2} \left( P \frac{D_1}{|D_1|} P
\right) e^{(- i \theta (J^2) \epsilon_2)/2},
\end{eqnarray} 
and
\begin{equation} 
\gamma = P \gamma_2 P.
\end{equation}	
They are all zero on $W$ and leave its orthogonal complement $V$
invariant. They have the following additional fundamental properties
on $V$: \\ 
(i) They are self-adjoint: 
\begin{equation} 
D^{\dagger} = D, F^\dagger = F, \gamma^\dagger = \gamma.
\label{sa}
\end{equation}
(ii)  $F$ and $\gamma$ square to ${\bf 1}$ on $V$: 
\begin{equation} 
F^2 = \gamma^2 = P.
\label{squareone}
\end{equation}
(iii) $\gamma$ anticommutes with $F$ and $D$:
\begin{equation}
\{\gamma, F\} =\{\gamma, D \} =0.
\label{anticom}
\end{equation} 
So $\gamma$ is a chirality operator on $V$. There is no fermion
doubling, and $D$ is an excellent approximation to ${\cal D}_\alpha$.

\section{The Projector $P$ and the Operator Algebra}
Our presentation above must have betrayed our intention to project all
to the subspace $V$. But we can consistently do so without disturbing
the mathematical formalism (even by small amounts) only if no operator
we deal with has matrix elements between $V$ and $W$. We must
therefore work only with operators commuting with $P$. 

This criterion is satisfied by $D, F$ and $\gamma$, but not by $a^L$
and $a^R$. As we cannot avoid their use, we correct them to commute
with $P$.

Let 
\begin{equation}
\Gamma(P) = 2P -{\bf 1}, \quad \Gamma(P)^2 = {\bf 1}.
\end{equation}
Then to any operator $\alpha$, we can associate another:
\begin{equation}
P(\alpha) = \alpha + \frac{1}{2} \Gamma(P) [ \alpha, \Gamma(P)] =
\frac{1}{2}[\alpha + \Gamma(P) \alpha \Gamma(P)].
\end{equation}
It commutes with $\Gamma(P)$,
\begin{equation}
P(\alpha) \Gamma(P) = \Gamma(P) P(\alpha)
\label{PwithA}
\end{equation}
and its projections to $V$ and $W$ coincide with those of $\alpha$:
\begin{equation}
P P(\alpha) P  = P\alpha P, \quad ({\bf 1} -P)P(\alpha)({\bf 1}-P) =
({\bf 1} -P)\alpha({\bf 1}-P). 
\end{equation}
We can think of $\frac{1}{2} \Gamma(P) [ \alpha, \Gamma(P)]$ as the
connection canonically extending the action of $\alpha$ from $V\oplus
W$ to $V$ and $W$.

It is interesting that 
\begin{eqnarray}  
P(\alpha)P(\beta) - P(\alpha \beta) &=& -\frac{1}{4}\Gamma(P) [ \alpha,
\Gamma(P)] \Gamma(P) [ \beta, \Gamma(P)]  \nonumber \\
&=& \frac{1}{4}[\alpha, \Gamma(P)][\beta, \Gamma(P)] \label{pdiff}
\end{eqnarray} 
where the right-hand side commutes with $\Gamma(P)$. It vanishes
certainly as $l \rightarrow \infty$. 

In future we will change operators like $a^{L,R}$ not commuting with
$P$ to their $P$-images. Let $P(A^{L,R})$ be the algebra generated by
$P(a^{L,R})$. From (\ref{pdiff}), we see that they will contain small
operators with no pre-image in $A^{L,R}$ and going to zero as $l
\rightarrow \infty$.

\section{Connections and Curvatures on $S_F^2$}
Connes' approach to gauge theories is based on spectral triples and
$K$-cycles
\cite{connes,landi,coquereaux,vargra,mssv,varilly}. Fredholm modules
and cyclic cohomology \cite{connes} seem better suited for fuzzy
physics, especially for maintaining continuum topological features
like instanton bounds and $\theta$-states. We have already seen this
in work on fuzzy monopoles \cite{bbiv}, and give further supporting
evidence in this paper. Later on, we will comment on the continuum
limit of this approach, its full details being reserved to future work
\cite{denjoe}.

Alternative approaches to gauge theories on the fuzzy sphere have also
been developed by Grosse and Pre\v{s}najder \cite{grklpr1} and
Klim\v{c}\'{\i}k \cite{klimcik}. 

In mathematics, Fredholm modules and cyclic cohomology have played a
central role in the $K$-theory of algebras for many years
\cite{connes}, while Rajeev and coworkers \cite{rajeev,ferraj} have
long appreciated their importance in the physics of large-$N$ gauge
theories. Here, we suggest their importance for fuzzy physics too.

We first introduce the concept of forms and then go on to formulate
gauge theories. All that we now do can be restricted to $V$. They are
also suitable for adaptation to general fuzzy spaces. However, the
treatment of instantons and monopoles is being postponed till Section
6.

\subsection{Forms}
These are constructed using $F, \gamma$ and $P(A^L)$. For this work,
in addition to (\ref{sa}, \ref{squareone}, \ref{anticom}), it is
important that
\begin{equation}
[\gamma, P(a^L)]=[\gamma, a^L +\frac{1}{2}\Gamma(P)[a^L,
\Gamma(P)]]=0.  
\end{equation}
This result is implied by the fact that $\gamma$ commutes with both
$a^L$ and $P$. 
 
Forms are linear spans of elements 
\begin{equation}
\omega_\lambda=P(a_0^L)[F, P(a_1^L)]\ldots[F, P(a_\lambda^L)] 
\label{form}
\end{equation}
for all $\lambda$. Their product can also be written as linear
combinations of terms like (\ref{form}) using properties of
derivations. (For instance, $[F, P(a^L)]P(b^L) = [F,
P(a^L)P(b^L)]-P(a^L)[F, P(b^L)]$).

Forms are ${\mathbb Z}_2$-graded by $\gamma$: those $\omega_\lambda$
with $\lambda$ even commute with $\gamma$ and give us the space of
even forms $\Omega^0[P(A^L)]$ while those $\omega_\lambda$ with
$\lambda$ odd anticommute with $\gamma$ and give us the space of odd
forms $\Omega^1[P(A^L)]$.

We can assign degrees $\partial \omega_\lambda$ to $\omega_\lambda$
(=0 or 1 if $\lambda=0$ or 1) which is additive (mod 2). The degree of
$\omega_\lambda \omega_\mu$ is $\lambda+\mu$ (mod 2).

There is a derivation $d$ between $\Omega^0[P(A^L)]$ and
$\Omega^1[P(A^L)]$ which squares to zero. It is given by the graded
commutator with $F$:
\begin{eqnarray}
d\omega_\lambda &=& F\omega_\lambda - (-1)^{\partial \omega_\lambda}
\omega_\lambda F \equiv 
\left[ F, \omega_\lambda \right\}, \\
d^2 &=& 0.
\end{eqnarray}
The operator $d$ resembles the differential in manifold theory and
leads to a homology theory for $P(A^L)$.

\subsection{Connections and Curvatures}
Connections $\nabla(\omega)$ and curvatures ${\cal F}(\omega)$ depend
on ``Lie-algebra valued one-forms'' $\omega$. For $U(M)$ gauge
theories, if $-iT(\alpha)$ ($\alpha = 0, 1, \ldots M^2-1$) are $M \times
M$ antihermitian matrices spanning the Lie algebra $\underline{U(M)}$
of $U(M)$, then
\begin{eqnarray}
\omega &=& T(\alpha) P(\omega^\alpha_1), \\
P(\omega^\alpha_1) &=& \sum_j P(a^{L,j}_\alpha)[F, P(b^{L,j}_\alpha)], \\
P(\omega^\alpha_1) &=& P(\omega^\alpha_1)^*, \quad P(a^{L,j}_\alpha),\;
P(b^{L,j}_\alpha) \in P(A^L).
\end{eqnarray}
$\omega$ is a linear operator on
\begin{equation}
A^M \otimes {\mathbb C}^2 = \{|\xi = (\xi_{\lambda j}) \rangle,
\lambda = 1, \ldots M; j=1,2, \, \xi_{\lambda j} \in A \}.
\label{intspace} 
\end{equation}
Here, $\lambda$ is the ``internal symmetry'' index on which $T(\alpha)$
act and $j$ is the spin index. The scalar product for $A^M \otimes
{\mathbb C}^2$ is the evident one:
\begin{equation}
\langle \eta| \xi \rangle = Tr \sum_{\lambda, j} \eta^*_{\lambda j}
\xi_{\lambda j}.
\end{equation}

$\nabla(\omega)$ and ${\cal F}(\omega)$ are the following operators:
\begin{eqnarray}
\nabla(\omega) &=& d + \omega, \\
{\cal F}(\omega) &=& d \omega + \omega^2 \equiv \{F, \omega\} +
\omega^2. 
\end{eqnarray}
[Both act on $P(A^L)^M \otimes {\mathbb C}^2 \equiv P(A^L)^{M+2}$,
with $d$ being the derivation. ${\cal F}(\omega)$ is also a linear
operator on $A^M \otimes {\mathbb C}^2$.]

The gauged version of $F$ on $A^M \otimes {\mathbb C}^2$ is this:
\begin{equation} 
\text{Gauged version of $F$} = F + \omega.
\end{equation} 

The $U(M)$ gauge group consists of $M \times M$ unitary matrices $u$
with $u_{ij} \in P(A^L)$. They act on $\nabla(\omega)$ and ${\cal
F}(\omega)$ in the usual way:
\begin{eqnarray}
\nabla(\omega) &\rightarrow& u \nabla(\omega) u^{\dagger}, \\
{\cal F}(\omega) &\rightarrow& u {\cal F}(\omega) u^{\dagger}.
\end{eqnarray}

We can describe gauge theories for a subgroup $G \subset U(M)$ by
restricting $\omega$. That is, if $\tilde{T}(\alpha)$ span the Lie
algebra $\underline{G}$ of $G$, we can decide to consider only $\omega
= \tilde{T}(\alpha)P(\omega^\alpha_1)$. (But that would not guarantee
that ${\cal F}(\omega)$ has an expansion containing only
$\tilde{T}(\alpha)$ since $P(\omega^\alpha_1)$ and $P(\omega^\beta_1)$
may not anticommute in the term $\omega^2 =
\tilde{T}(\alpha)\tilde{T}(\beta) P(\omega^\alpha_1)
P(\omega^\beta_1)$.) In the same vein, when considering covariant
derivative in a representation $\Gamma$ of $U(M)$, we change
$T(\alpha)$ to its representative in the Lie algebra
$\underline{\Gamma}$ of $\Gamma$.

\section{The Actions and Quantization}
\subsection{Critical Dimensions}
In Connes' approach, the Euclidean actions for free massless scalars,
spinors and gauge fields on a manifold of dimension $n$ and Dirac
operator ${\cal D}$ are respectively,
\begin{eqnarray}
{\cal S}(\Phi) &=& \text{constant}\;Tr^{+} \left[\frac{1}{|{\cal D}|^n}
\left([{\cal D}, \Phi]^{\dagger}[{\cal D}, \Phi] \right)\right],
\label{CscalarS}\\ 
{\cal S}(\Psi) &=& \text{constant}\;Tr^{+} \left[\frac{1}{|{\cal D}|^n}
\Psi^{\dagger} {\cal D} \Psi \right], \label{CspinorS}\\
{\cal S}(\hat{\omega}) &=& \text{constant}\;Tr^{+}
\left[\frac{1}{|{\cal D}|^n} \left( [{\cal D}, \hat{\omega}] +
\hat{\omega}^2\right)^{\dagger} \left( [{\cal D}, \hat{\omega}] +
\hat{\omega}^2\right) \right]. \label{CgaugeS}
\end{eqnarray}
Here $Tr^{+}$ is Dixmier trace, $n$ is spacetime dimension, and $\Phi,
\Psi$ and $\hat{\omega}$ are scalar, spinor and gauge fields
respectively. [Issues involving ``junk forms'' are being ignored.]

Under the scaling transformation ${\cal D} \rightarrow \lambda {\cal
D}$, the response of $\omega$ is $\omega \rightarrow \lambda
\omega$. Hence under ${\cal D} \rightarrow \lambda {\cal D}$, 
\begin{eqnarray}
{\cal S}(\Phi) &\rightarrow& \lambda^{2-n} {\cal S}(\Phi),
\label{scresp}\\ 
{\cal S}(\Psi) &\rightarrow& \lambda^{1-n} {\cal S}(\Psi),
\label{spresp}\\ 
{\cal S}(\hat{\omega}) &\rightarrow& \lambda^{4-n} {\cal
S}(\hat{\omega}). \label{gauresp} 
\end{eqnarray}
Gauging ${\cal D}$ does not affect (\ref{scresp}) and
(\ref{spresp}).  

The critical dimensions where actions are scale-invariant are thus \\
$n=2$ for $\Phi$, $n=1$ for $\Psi$ and $n=4$ for $\hat{\omega}$.

We first propose actions for fuzzy scalars and gauge fields in their
critical dimensions, $n=1$ being outside our modeling scope. 

\subsection{$n=2$ Fuzzy Massless Scalar Fields} 
A fuzzy scalar field $\phi$ for $n=2$ is a polynomial in
$P(a^L)$. If it has internal degrees of freedom, it is a vector with
each component $\phi_\rho$ being such a polynomial.

Our (Euclidean) action for a zero mass ``non-interacting'' fuzzy
scalar field $\phi$ is
\begin{equation}
S(\phi) = f^2 Tr P[d P(\phi)]^{\dagger} [d P(\phi)] \label{fscalarS}
\end{equation}
where the internal index $\rho$, if any, is summed within the
trace. It can be gauged in an evident manner by replacing $d$ by
$\nabla(\omega)$. It is scale invariant just like (\ref{CscalarS}).

In ref \cite{bbiv}, the analogue of this action for fuzzy
$\sigma$-models was proposed and its Belavin-Polyakov bound
\cite{belpol} was discussed. As suggested there, we conjecture that as
$l \rightarrow \infty$, $f^2$ can be scaled in such a way that
$S(\phi)$ approaches ${\cal S}(\Phi)$. Work on this matter is in
progress \cite{denjoe}.

\subsection{$n=4$ Fuzzy Gauge Field}
The formalism of cyclic cohomology and gauge theories depends only on
the knowledge of suitable $d$ and chirality operators. For $n=4$,
there are two fuzzy spaces that are susceptible to our analysis,
namely $S_F^2 \times S_F^2$ and $({\mathbb C}{\mathbb P}^2)_F$. The
former is enough for illustration. The algebra for that space is $A
\otimes_{\mathbb C} A$ while its $d$ and chirality operators are
$\left[ 1/\sqrt{2} (F \otimes_{\mathbb C} {\bf 1} + \gamma
\otimes_{\mathbb C} F),.\right\}$ and $\gamma \otimes \gamma$
respectively ($F$ and $\gamma$ being those of $S_F^2$.)

Our proposed action is the evident one:
\begin{equation}
S(\omega) = \frac{1}{2 e^2} Tr P\left\{ {\cal F}(\omega) {\cal F}(\omega)
\right\} \label{fgaugeS}. 
\end{equation}
It is accompanied by a conjecture like that for $S[\phi]$ about its
$l \rightarrow \infty$ limit. It is scale invariant like ${\cal
S}(\hat{\omega})$ of $n=4$.

For $({\mathbb C}{\mathbb P}^2)_F$ as well, our action looks the same as
(\ref{fgaugeS}). But we have yet to tell what $d$ and the chirality
operator are, a task postponed to later work.

\subsection{Away from Criticality}
Our guide for the choice of actions for any $n$ continues to be
scaling properties and gauge invariance. Also in our formulas for
general $n$, the precise definitions of $P, D, F$ and $D_1$ require
future elucidation.

{\it (a) Scalars} \\
The action for $\phi$ for any $n$ is suggested by (\ref{scresp}) to be
\begin{equation} 
S(\phi) = f^2 Tr P |D|^{2-n} [\nabla (\omega) P(\phi)]^{\dagger}
[\nabla(\omega) P(\phi)]
\end{equation} 
[For $n=2, |D|$ is invertible on $V$.]

{\it (b) The field $g$} \\
The formulation of gauge invariant actions for fuzzy spinor and gauge
fields requires the introduction of a matrix `field' $g$. Its
components $g_{ab}$ are polynomials in $P(a^L)$. It is unitary and
commutes with $P$:
\begin{eqnarray} 
g^{\dagger}g &=& {\bf 1}, \\
gP &=& Pg.
\end{eqnarray}   
The index $a$ carries the action of the gauge transformations $u$:
\begin{equation} 
u: g \rightarrow ug
\end{equation} 

{\it (c) Spinors} \\ A fuzzy spinor $\psi$ is an element of the
Hilbert space on which operators like $\gamma, D$ and elements of
$A^{L,R}$ can act. In the presence of internal symmetry, it has
components $\psi_{\lambda,j}$ (c.f. \ref{intspace}). The gauged action
is suggested by (\ref{spresp}):
\begin{equation}
S(\psi,g) = \kappa \langle \psi|Pg \left( \frac{1}{|D|^{(n-1)/2}}
\right) g^\dagger (F+\omega)g \left(\frac{1}{|D|^{(n-1)/2}}
\right)g^\dagger P|\psi \rangle. \label{spinaction} 
\end{equation}
It depends on both $\psi$ and $g$.

We can now define a new Dirac field 
\begin{equation}
|\chi\rangle = g \left(\frac{1}{|D|^{(n-1)/2}}\right)g^\dagger | \psi
\rangle 
\end{equation}
which transforms in the same way as $|\psi \rangle$ under the gauge
group,
\begin{equation}
u: |\chi \rangle \rightarrow u |\chi \rangle ,
\end{equation} 
but which scales differently:
\begin{equation}
|\chi \rangle \rightarrow \lambda^{(1-n)/2} |\chi \rangle \quad
\text{under} \quad D \rightarrow \lambda D. \label{chiresp}
\end{equation}
The spinor action is thus 
\begin{equation}
S(\chi) = \kappa  \langle \chi |P[F+\omega]P| \chi \rangle 
\end{equation}
where $\chi$ scales as in (\ref{chiresp}). (We assume as for $n=2$
that $[|D|, P]=0$.)

{\it (d) Gauge Fields} \\ The $n\neq4$ gauge field action, like
(\ref{spinaction}), depends on both $\omega$ and $g$ and reads
\begin{equation}
S(\omega,g) = \frac{1}{2e^2} Tr P\left[ g |D|^{4-n} g^\dagger
\right]{\cal F}(\omega)^2. \label{gaugeaction}
\end{equation}
compatibly with (\ref{gauresp}).

There is a certain freedom in the choice of gauge field action. Any
one of the following actions are a priori equally acceptable:
\begin{equation}
S_a(\omega,g) = \frac{1}{2e^2} Tr P\left[ g |D|^{a} g^\dagger \right]
{\cal F}(\omega) \left[ g |D|^{b} g^\dagger \right] {\cal F}(\omega),
\quad a+b = 4-n.
\label{gactiona}
\end{equation}

{\it (e) A Remark} \\ In the continuum limit, if $Tr$ goes over to
$Tr^+$ as we conjecture, we can cancel $g$ with $g^\dagger$ in
(\ref{spinaction}), (\ref{gaugeaction}) and (\ref{gactiona})
\cite{connes,vargra,varilly} and the dependence on $g$ disappears.

{\it (f) Mass and Interaction Terms} \\
Mass and interaction terms can also be introduced with guidance from
continuum scaling properties and from gauge invariance. We omit the
simple details.

{\it (g) Quantization} \\ Quantization can be done using functional
integration. We can for example try to expand the fields in normal
modes and integrate $\exp{(-\text{action})}$ over the coefficients in
the mode expansions to define the partition function. This method is
especially useful for the Dirac field for which the normal modes are
given by an orthonormal basis of the Hilbert space and the
coefficients of the expansion are Grassmann numbers.

\section{On Twisted Bundles and Fuzzy Physics} 
In the continuum, instantons are particular connection fields $\omega$
on certain twisted bundles over the base manifold ${\cal M}$. On
$S^2$, they are monopole bundles, on $S^4$ or ${\mathbb C}{\mathbb
P}^2$, they can be $SU(2)$ bundles. For such reasons, we may
henceforth refer to monopoles also as instantons.

In algebraic $K$-theory
\cite{connes,landi,coquereaux,vargra,mssv,varilly,landi:98}, it is
well-known that these bundles are associated with projectors ${\cal
P}$.  ${\cal P}$ is a matrix of some dimension $M$ with ${\cal P}_{ij}
\in {\cal A} \equiv {\cal C}^{\infty}({\cal M})$, ${\cal P}^2 = {\cal
P} = {\cal P}^{\dagger}$. The physical meaning of ${\cal P}$ is the
following. Let ${\cal A}^M = {\cal A} \otimes {\mathbb C}^M = \{ \xi
=(\xi_1, \xi_2 \ldots \xi_M):\xi_i \in {\cal A} \}$. Then ${\cal PA}^M
= \{ {\cal P}\xi: \xi \in {\cal A}^M\}$ consists of smooth sections
(or wave functions) ${\cal P} \xi$ of a vector bundle over ${\cal
M}$. For suitable choices of ${\cal P}$, we get monopole or instanton
vector bundles. These projectors are known \cite{mssv,landi:98} and
those for monopoles have been reproduced in \cite{bbiv}.

The projectors $p^{(\pm N)}$ for fuzzy monopoles of charge $\pm N$
have also been found in \cite{bbiv}. They act on $A^{2^N} = \{ \xi$
with components $\xi_{b_1 \ldots b_N} \in A, b_i \in \{1,2\} \}$.
Let $\vec{\tau}^{(i)}$ ($i=1, 2, \ldots N$) be commuting Pauli
matrices. $\vec{\tau}^{(i)}$ has the normal action on the index $b_i$
and does not affect $b_j$ ($j \neq i$). Then $\vec{K} = \vec{L^L} +
\sum_i \vec{\tau}^{(i)}/2$ generates the $SU(2)$ Lie algebra and
$p^{(N)}$ ($p^{(-N)}$) is the projector to the maximum (minimum)
angular momentum $k_{\text{max}}=l+N/2$
($k_{\text{min}}=l-N/2$). [$\vec{K}^2 p^{(N)} =
k_{\text{max}}(k_{\text{max}}+1) p^{(N)}$, $\vec{K}^2 p^{(-N)} =
k_{\text{min}}(k_{\text{min}}+1) p^{(-N)}$.] Fuzzy analogues of
monopole wave functions are $p^{(\pm N)} A^{2^N}$. Explicit
expressions for $p^{(\pm N)}$ may be found in \cite{bbiv}.

When spin is included, we must enhance $p^{(\pm N)} A^{2^N}$ to
$p^{(\pm N)} A^{2^N} \otimes {\mathbb C}^2 = p^{(\pm N)} A^{2^{N+1}} =
\{\xi$ with components $\xi_{b_1 \ldots b_N, j} \in A: b_i, j \in
\{1,2 \} \}$.

As for four-dimensional instantons, we do not know their projectors
even for $S_F^2 \times S_F^2$, so we shall just assume their existence
in what follows. That is enough for the presentation because of its
generality.

The discussion below focuses on $S_F^2$, but one can readily see how
to go beyond this space, once the basic ingredients become available.

\subsection{Cyclic Cohomology of Twisted Sectors}
All the complications resolved here are caused by the need to project
out a subspace of $A^{2^{N+1}}$. It is the analogue of the subspace
projected out by $P$ for $N=0$. In its absence, for example in the
continuum, there is a canonical way to extend cyclic cohomology to
twisted bundles. It is also due to Connes \cite{connes}.

The material being explained now has been partially reported in
\cite{ongoing}. It is not essential reading in all its details for
what follows once it is accepted that a certain subspace of
$A^{2^{N+1}}$ can be consistently projected out.

In the $N=0$ sector, the projector $P$ cuts out the subspace $W$ of
$A^2$. The function of the map $\alpha \rightarrow P(\alpha)$ of
operators is to make them compatible with the splitting $A^2 = V
\oplus W$.

When we pass to $p^{(\pm N)} A^{2^N}$ and thence to $p^{(\pm N)}
A^{2^{N+1}}$ by including spin, the subspace to be projected out is
{\it not} determined by $P$ if $N \neq 0$, as we shall see
below. Rather, we can explain it as follows: Let $\vec{J} = \vec{K} -
\vec{L^R} + \vec{\sigma}/2$ be the ``total angular momentum''. Calling
$\vec{J}$ by this name is appropriate as it becomes (\ref{tamo}) for
$N=0$ and displays the known ``spin-isospin mixing''
\cite{jacreb,hastho} for $N \neq 0$. The maximum of $\vec{J}^2$ on
$p^{(\pm N)} A^{2^{N+1}}$ is $J_{\text{max}}(J_{\text{max}} +1),
J_{\text{max}} = (l \pm N/2)+l+1/2 =2l \pm N/2 +1/2$. [We assume that
$2l \geq (N-1)/2$.] The vectors to be projected out are those with
total angular momentum $J_{\text{max}}$. If ${\cal J}^{(\pm N)}$ are
the corresponding projectors [with ${\cal J}^{(0)} = P$], the twisted 
space we work with is thus ${\cal J}^{(\pm N)}p^{(\pm N)}
A^{2^{N+1}}$. Since $p^{(\pm N)}$ commute with $\vec{J}$ and hence with
${\cal J}^{(\pm N)}$, $Q^{(\pm N)} = {\cal J}^{(\pm N)}p^{(\pm
N)}$ are also projectors. 

There is no degeneracy for angular momentum $J_{\text{max}}$ in
$p^{(\pm N)} A^{2^{N+1}}$. That is because there is only one way to
couple $l \pm N/2, l$ and $1/2$ to $J_{\text{max}}$. The space $({\bf
1} - {\cal J}^{(\pm N)})p^{(\pm N)} A^{2^{N+1}}$ is thus of dimension
$2 J_{\text{max}} + 1$. We want to get rid of this subspace.

The operators $T = D, F$ or $\gamma$ of Section 2 are zero on $({\bf
1} - P)A^2$ where $P$ cuts out states of angular momentum
$2l+1/2$. There is no degeneracy for this angular momentum in
$A^2$. $T$ and $P$ extend canonically to $A^2 \otimes {\mathbb
C}^{2^N} (\equiv A^{2^{N+1}})$ as $T \otimes {\bf 1}$ and $P \otimes
{\bf 1}$. Let us call them once more as $T$ and $P$. $T$ and $P$
commute with $\vec{J}$ and hence with ${\cal J}^{(\pm N)}$. There is
only one way to couple $N$ ``isospin'' $1/2$'s to $(2l + 1/2)$ to get
$J_{\text{max}}$ so that $({\bf 1} - {\cal J}^{(\pm N)})({\bf 1} - P)
A^{2^{N+1}}$ is also of dimension $2 J_{\text{max}}+1$. And $T$ is
zero on this subspace.

The projectors $({\bf 1} - {\cal J}^{(\pm N)})p^{(\pm N)}$ and $({\bf
1} - {\cal J}^{(\pm N)})({\bf 1} - P)$ being of the same rank, there
exists a unitary operator $U$ on $A^{2^{N+1}}$ transforming one to the
other:
\begin{equation} 
({\bf 1} - {\cal J}^{(\pm N)})p^{(\pm N)} = U({\bf 1} - {\cal J}^{(\pm
N)})({\bf 1} - P)U^{-1}. 
\label{Udef}
\end{equation} 
If we transport $T$ by $U$,
\begin{equation} 
T' = UTU^{-1},
\end{equation} 
then $T'=D', F'$ or $\gamma'$ vanishes on $({\bf 1} - {\cal J}^{(\pm
N)}) p^{(\pm N)} A^{2^{N+1}}$. On its orthogonal complement $[{\bf 1}
- ({\bf 1} - {\cal J}^{(\pm N)})p^{(\pm N)}] A^{2^{N+1}}$, invariant
under $T'$, $F'$ and $\gamma'$ square to unity and $\gamma'$
anticommutes with $D'$ and $F'$, just as we want. What replaces
$P$ now is not ${\cal J}^{(\pm N)}$, but rather
\begin{eqnarray} 
P^{(\pm N)} &=& [{\bf 1}-({\bf 1}-{\cal J}^{(\pm N)})p^{(\pm N)}], \\ 
P^{(0)} &=& P.
\end{eqnarray} 

As $l \rightarrow \infty$, $J_{\text{max}}$ becomes dominated by $2l$
and so we have the freedom to let $U$ approach ${\bf 1}$. That is, no
$U$ is needed in the continuum limit. 

Since $p^{(\pm N)}$ define different topological sectors, we also have
the freedom to choose different $U$'s for these sectors. But as both
these sectors come from $A^{2^{N+1}}$, it is convenient to find a
single $U$ valid for both.

Total angular momentum $2l+1/2+N/2$ has no multiplicity in
$A^{2^{N+1}}$. As both $({\bf 1} - {\cal J}^{(N)})({\bf 1} - P)
A^{2^{N+1}}$ and $({\bf 1} - {\cal J}^{(N)})p^{(N)} A^{2^{N+1}}$ have
this angular momentum, we have that 
\begin{equation} 
({\bf 1} - {\cal J}^{(N)})({\bf 1} - P) A^{2^{N+1}} = 
({\bf 1} - {\cal J}^{(N)})p^{(N)} A^{2^{N+1}}.
\label{same}
\end{equation}
So we choose
\begin{equation} 
U={\bf 1} \quad \text{on} \quad ({\bf 1} - {\cal J}^{(N)})({\bf 1}-P)
A^{2^{N+1}}. 
\label{Uchoice}
\end{equation}

Next in accordance with (\ref{Udef}), we set 
\begin{equation} 
U({\bf 1} - {\cal J}^{(-N)})({\bf 1} - P) A^{2^{N+1}} =
({\bf 1} - {\cal J}^{(-N)})p^{(-N)} A^{2^{N+1}}.
\label{UonminusN}
\end{equation}
We also demand that 
\begin{equation}
[U, J_i]=0.
\label{rotinv}
\end{equation}
That fixes $U$ upto a phase on the subspace $({\bf 1} - {\cal
J}^{(-N)})({\bf 1} - P) A^{2^{N+1}}$.

We saw in \cite{bbiv} that $(1 \pm \gamma)/2$ are projectors for
combining $-\vec{L}^R$ and $\vec{\sigma}/2$ to give angular momenta $l
\pm 1/2$. So $\gamma =+1$ on all the subspaces $({\bf 1} - {\cal
J}^{(\pm N)})({\bf 1}-P) A^{2^{N+1}}$, $({\bf 1} - {\cal J}^{(\pm
N)})(p^{(\pm N)}) A^{2^{N+1}}$. Also, $(\frac{\sum
\vec{\tau}^{(i)}}{2})^2$ is $\frac{N}{2}(\frac{N}{2}+1)$ on these
subspaces. It follows that (\ref{Uchoice}), (\ref{UonminusN}) and
(\ref{rotinv}) are compatible with a $U$ commuting with $J_i$,
$\gamma$, $(-\vec{L}^R +\vec{\sigma}/2)^2$ and $(\frac{\sum
\vec{\tau}^{(i)}}{2})^2$. We now outline an extension of $U$ to all of
$A^{2^{N+1}}$ consistently with rotational invariance (\ref{rotinv})
and
\begin{equation} 
\left[U, \left( -\vec{L}^R+\frac{\vec{\sigma}}{2} \right)^2 \right] = 
\left[U, \left(\frac{\sum \vec{\tau}^{(i)}}{2}\right)^2 \right]= 0.
\end{equation}
An important consequence is that
\begin{equation} 
\gamma' = \gamma
\end{equation}
so that 
\begin{equation}
[\gamma' , p^{(\pm N)}] = [\gamma' , J_i] = [\gamma' , {\cal J}^{(\pm
N)}] = 0.
\label{chiprop}
\end{equation}

One way to specify $U$ more fully is as follows. Let 
\begin{equation} 
A^{2^{N+1}} = X \oplus X^{\perp} = X' \oplus {X'}^{\perp}
\end{equation}
be orthogonal decompositions where 
\begin{eqnarray}	
X &=& ({\bf 1}-{\cal J}^{(N)})({\bf 1} - P)A^{2^{N+1}} \oplus ({\bf
1}-{\cal J}^{(-N)})({\bf 1} - P)A^{2^{N+1}}, \\
X' &=& UX = ({\bf 1}-{\cal J}^{(N)})p^{(N)}A^{2^{N+1}} \oplus ({\bf
1}-{\cal J}^{(-N)}) p^{(-N)} A^{2^{N+1}}.
\end{eqnarray}
Both $X$ and $X'$ are invariant under the self-adjoint operators
\begin{equation} 
J_i, \left( -\vec{L}^R + \frac{\vec{\sigma}}{2} \right)^2 \quad 
\text{and} \quad \left( \sum \frac{\vec{\tau}^{(i)}}{2} \right)^2. 
\nonumber 
\end{equation}
Therefore, the same is the case with $X^{\perp}$ and
${X'}^{\perp}$. We can extend $U$ to a map $X^{\perp} \rightarrow
{X'}^{\perp}$ which commutes with the above operators. There would
still be uncertainties about choosing $U$ requiring further
conventions for elimination. 

The analogue of the $N=0$ map $\alpha \rightarrow P(\alpha)$ is just
$\alpha \rightarrow P^{(\pm N)}(\alpha)$.

We can now reproduce Sections 4 and 5 for any $N$ using $T'$ and
$P^{(\pm N)}(A^L)$. 

Although $T'$ and forms are operators on $P^{(\pm N)}A^{2^{N+1}}$,
that is not the space of sections for the twisted bundles. The latter
is, rather, 
\begin{equation}
Q^{(\pm N)} A^{2^{N+1}} = p^{(\pm N)}{\cal J}^{(\pm N)}A^{2^{N+1}} =
p^{(\pm N)}P^{(\pm N)} A^{2^{N+1}}.  
\end{equation} 
It is not an invariant subspace for $D'$ and $F'$ unless they are
projected, or corrected by connections. We shall do that
below. However, chirality $\gamma'$ is well-defined on twisted sections
because of (\ref{chiprop}).

For notational simplicity, we now permanently rename $T'$, $P^{(\pm
N)}$ and $Q^{(\pm N)}$ as follows:
\begin{eqnarray}	
D', F', \gamma' &\rightarrow& D, F, \gamma, \\
p^{(\pm N)} &\rightarrow& p, \\
P^{(\pm N)} &\rightarrow& P, \\
Q^{(\pm N)} &\rightarrow& Q.
\end{eqnarray} 
$\gamma'$ in any case is $\gamma$.

\section{Fuzzy Instantons, Topological Susceptibility, $\theta$-Term}
We will now deal with a generic space ${\cal M}_F$ and let $A$ stand
for its algebra. We will also assume that their cyclic cohomology and
instanton wave functions are described by operators $D, F, \gamma$ and
projectors $p, P$ and $Q=pP$ just as for $S_F^2$. 

A generic operator $P(\alpha)$ will not commute with $p$. It must
be changed to $pP(\alpha)p$.

It is often more convenient, as with $P(\alpha)$, to work with $p
\cdot P(\alpha) := p(P(\alpha))$:
\begin{eqnarray}	
p \cdot P(\alpha) &\equiv& pP(\alpha)p + ({\bf 1}-p) P(\alpha) 
({\bf 1}-p)  \nonumber \\  
&= & P(\alpha) + \frac{1}{2} \Gamma(p)[P(\alpha), \Gamma(p)], \\
\Gamma(p) &=& 2p - 1.
\end{eqnarray} 

The modification of $d$ is accordingly
\begin{equation}
p(d) = d + \frac{1}{2} \Gamma(p) \left(d \Gamma(p) \right).
\end{equation}
It contains the minimum irreducible gauge term $\Gamma(p) (d
\Gamma(p))/2$. Here, there can be further fluctuations of the kind $p
\hat{\omega}_1 p$ [we can also use $p \cdot \hat{\omega}_1$] so that
the general connection or covariant derivative is
\begin{eqnarray}
\nabla(\omega) &=& d + \frac{1}{2} \Gamma(p) \left(d \Gamma(p) \right)
 + p \hat{\omega}_1 p, \nonumber \\ 
&\equiv& d+\omega, \quad \hat{\omega}_1 =\text{a matrix of one-forms}.
\end{eqnarray}

The curvature can be read off now:
\begin{eqnarray} 	
\lefteqn{{\cal F}(\omega)=d \omega + \omega^2 = } \nonumber \\
& &d \left[ \frac{1}{2} \Gamma(p) \left(d \Gamma(p) \right) +
p \hat{\omega}_1 p
\right] + \left[\frac{1}{2} \Gamma(p) \left(d \Gamma(p) \right) +
p \hat{\omega}_1 p \right]^2.
\label{fcurv}
\end{eqnarray} 
Our action in the twisted sectors is also like (\ref{gaugeaction}):
\begin{equation}
S[\omega, g] = \frac{1}{2e^2} Tr Q\left( g|D|^{4-n} g^{\dagger}\right)
{\cal F} (\omega)^2. 
\end{equation}

For $n=4$, let 
\begin{equation} 
S[\omega, g] = \frac{1}{2e^2} \hat{S}(\omega).
\end{equation} 
$\hat{S}[\omega]$ has the topological (${\mathbb Z}$-valued)
lower bound $N$ for $p=p^{(\pm N)}$. It is like the bound on continuum
action saturated by instantons. That is enough to identify fuzzy
instantons, topological susceptibility and $\theta$-term.

The bound follows from the inequality
\begin{equation}
\left( \frac{1\pm \gamma}{2} Q {\cal F} (\omega) \right) \left(
\frac{1\pm \gamma}{2} Q {\cal F} (\omega) \right)^{\dagger} \geq 0
\end{equation}
by tracing, where $\geq$ indicates a nonnegative operator:
\begin{equation}
Tr Q {\cal F}(\omega)^2 \geq |Tr \gamma Q {\cal F}(\omega)^2|.
\end{equation}

In the next section, we show that 
\begin{equation} 
Tr \gamma Q {\cal F}(\omega)^2 
\label{index}
\end{equation}  
is independent of $\hat{\omega}_1$ and is the index of the operator
\begin{equation}
\frac{1-\gamma}{2}pFp\frac{1+\gamma}{2} \equiv F_+:
\frac{1+\gamma}{2}QA^{K+s} \rightarrow \frac{1-\gamma}{2}QA^{K+s}.
\end{equation}
This index is the difference of the dimensions of the subspaces of
$QA^{K+s}$ with $\gamma=+1$ and $\gamma=-1$. 

The bound is saturated if 
\begin{equation}
\frac{1\pm \gamma}{2}Q{\cal F}(\omega) = 0 \label{satb}.
\end{equation}
We can thus regard (\ref{satb}) as defining fuzzy instantons and
anti-instantons.

We can go a step further and propose a topological susceptibility. Let
$tr$ (with a lower case $t$) indicate trace over spin and internal
indices (the latter labeling components of $A^K$). Then fuzzy
topological susceptibility is
\begin{equation}
tr \gamma Q {\cal F} (\omega)^2. \label{fsus}
\end{equation}
Dimensional reasons indicate that it is to be identified with
$d(vol){\cal G}^*{\cal G}$ if ${\cal G}$ is the continuum curvature,
$n$ is 4 and $d(vol)$ the volume form.

The proposal (\ref{fsus}) for topological susceptibility is valid for
any $p$, so also in the $N=0$ sector ($p={\bf 1}$).

The $\theta$-term in the fuzzy action ``density'' is proportional to
(\ref{fsus}), being just
\begin{equation}
i \theta tr \gamma Q {\cal F} (\omega)^2.
\end{equation}

Electrodynamics on $S^2$ has the term $\theta \int {\cal G}$. Its
analogue here is 
\begin{equation}
i \theta tr \gamma Q {\cal F} (\omega) \label{ftheta}.
\end{equation}
Gauge theories for spacetime dimension $2n$ usually admit the
$\theta$-terms
\begin{equation}
i \theta \int tr(\underbrace{{\cal G} \wedge {\cal G} \wedge \ldots
{\cal G}}_{\text{$n$ factors}})
\end{equation}
Their analogues too exist for fuzzy spaces, being
\begin{equation}
i \theta tr \gamma Q {\cal F} (\omega)^n \label{fthetan}
\end{equation}
(\ref{fthetan}) is independent of $\hat{\omega}_1$ for all $n \geq 1$
and is the index of $F_+$. It is thus the same for any $n \in \{1, 2,
\cdots \}$.

\section{Axial Anomaly}
We show three connected results for $S_F^2$ in this section. \\ (a) On
the space $QA^{K+2}$, the Dirac operator $QDQ$ as also $QFQ$ have
exactly $N$ zero modes of chirality $\gamma=+1$ if $p$ is $p^{(N)}$
(and chirality $\gamma=-1$ if $p$ is $p^{(-N)}$). \\ (b) $\pm N$,
which are the indices of $F_+$, are given by (\ref{index}) for
$\hat{\omega}_1=0$. \\ (c) The expression (\ref{index}) is independent
of $\hat{\omega}_1$.

In view of Fujikawa's argument \cite{fujikawa}, (a) shows how axial
anomaly appears (in its integrated form) in the presence of
instantons. (b) and (c) supply the missing arguments for the last
section. While the focus is on $S_F^2$, the methodology is not so
limited and can generalize to any fuzzy space.

\subsection{Instantons and Zero Modes}

We prove (a) in this subsection.

The operators $QDQ=pDp$, $QFQ=pFp$ and $\gamma$ commute with $\vec{J}
= \vec{K} - \vec{L}^R + \vec{\sigma}/2$. Spaces with fixed $\vec{J}^2,
J_3$ are thus invariant under these operators.

For fixed $\vec{J}^2, J_3$, let $\Lambda$ be the eigenvalue of either
of the operators $T = pDp$ or $pFp$. As $\gamma$ anticommutes with $T$
and commutes with $\vec{J}$, for each eigenstate with $\Lambda \neq
0$, there is another with $-\Lambda$ and same $\vec{J}^2,
J_3$. Eigenvalues $\pm \Lambda \neq 0$ come in pairs with same
$\vec{J}^2, J_3$.

Consider for specificity $p=p^{(N)}$. Then $k_{\text{max}}=l+N/2$
gives the eigenspace $k_{\text{max}}(k_{\text{max}}+1)$ for
$\vec{K}^2$. Adding $-\vec{L}^R$ and then $\vec{\sigma}/2$ gives the
spectrum $J(J+1)$ of $\vec{J}^2$. We find $J=\{(l+N/2)+l, \cdots
(l+N/2)-l\} \pm 1/2$.

\begin{figure}[t]
\centerline{\epsfig{file=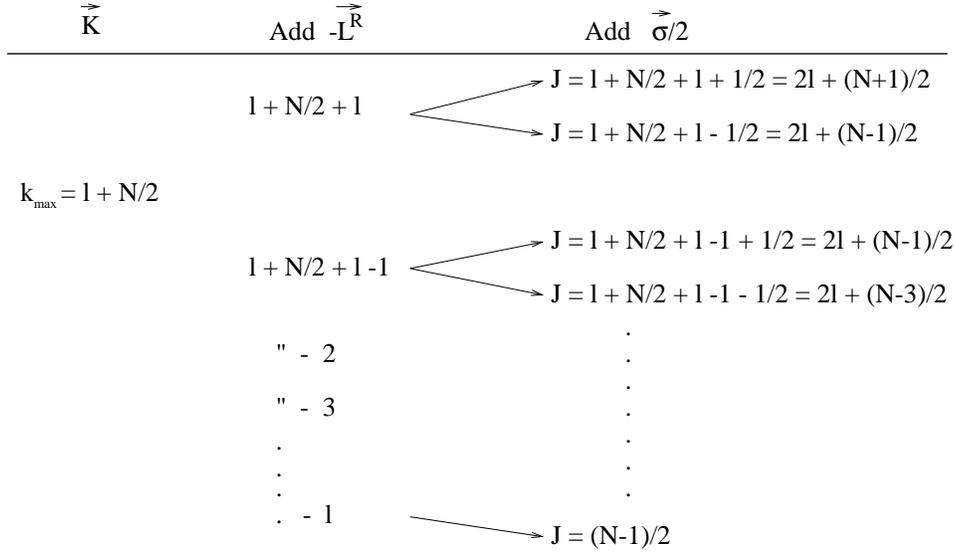,clip=5in,width=5in}}
\caption{Addition of Angular Momenta for $p=p^{(N)}$.}
\label{addition}
\end{figure}

Figure \ref{addition} shows how the addition of angular momenta is
working. This explanation is a bit of a repetition of what we did in
6.1.

All the $J$'s occur with multiplicity 2 except the topmost with $J =
J_{\text{max}} = 2l + (N+1)/2$ and the lowermost with $J =
J_{\text{min}} = (N-1)/2$. Vectors with $J_{\text{max}}$ and
$J_{\text{min}}$ are necessarily zero modes of $T$.

Vectors with $J_{\text{max}}$ are unphysical just like their $N=0$
siblings in Section 2.  Now $P$ projects out the unpaired state with
$J=J_{\text{max}}$. Hence the remaining zero modes after projection by
$pP$ have $J = J_{\text{min}}$. Their multiplicity is $N$ as
claimed. Also $-\vec{L}^R$ and $\sigma/2$ must combine to $l+1/2$ to
create $J_{\text{min}}$ as $l-1/2$ cannot combine with
$k_{\text{max}}=l+N/2$ and give $(N-1)/2$. The projector for getting
this $l+1/2$ is just $(1 + \gamma)/2$ \cite{bbiv}. Thus these zero
modes have left helicity as claimed.

Next let $p=p^{(-N)}$. Then $k_{\text{max}}$ is changed to
$k_{\text{min}} = l-N/2$ and the unpaired $J_{\text{max}} = 2l
-(N-1)/2$ is once more eliminated by $P$. All other $J$ are paired
except for $(l-1/2)-(l-N/2)=(N-1)/2$. These give the necessary zero
modes. There are $N$ of them. As the projector for $l-1/2$ is
$(1-\gamma)/2$ \cite{bbiv}, they are right-chiral. All these are as
claimed.

We have now established that $\frac{1-\gamma}{2} T \frac{1+\gamma}{2}$
has index $\pm N$ if $p=p^{(\pm N)}$. 

Let $\xi$ be a smooth perturbation of $T$ (for example due to
$\hat{\omega}_1$) changing it from $T(0)=T$ to $T(\xi)$ such that
$\gamma T(\xi) + T(\xi) \gamma =0$. We can say that if
$V_{\Lambda(\xi)}$ is the eigenspace of $T(\xi)$ for eigenvalue
$\Lambda(\xi) \neq 0$ $[\Lambda(0) = \Lambda]$, then $\gamma
V_{\Lambda(\xi)} = V_{-\Lambda(\xi)}$ is its eigenspace for opposite
eigenvalue and with same dimension. Also the multiplicities of states
with opposite chiralities in $V_{\Lambda(\xi)} \oplus
V_{-\Lambda(\xi)}$ are equal. $V_{\pm\Lambda(\xi)}$ are just the
deformations of $V_{\Lambda(0)}$ as $\xi$ is continuously turned on
and have dimensions independent of $\xi$. As the zero modes of $T$
have a unique chirality, it follows that they cannot become modes with
non-zero eigenvalues as $\xi$ is changed, lacking vectors with
opposite chirality to pair with. The index of $\frac{1-\gamma}{2} T
\frac{1+\gamma}{2}$ is hence stable under such perturbations.

\subsection{The Index and Curvature}
Let us prove (b) now.

The curvature ${\cal F}(\frac{1}{2}\Gamma(p)d\Gamma(p))$ for
$\hat{\omega}_1=0$ follows from (\ref{fcurv}):
\begin{equation}
{\cal F}(\frac{1}{2}\Gamma(p)d\Gamma(p))=(dp)(dp).
\end{equation}
We first show that 
\begin{equation}
Tr \gamma Q {\cal F}(\frac{1}{2}\Gamma(p)d\Gamma(p))^n = \pm N \quad
\text{for} \quad p=p^{(\pm N)} \quad \text{and for any $n$}. 
\end{equation}
Here $n= \{1, 2, \cdots \}$ and need not have any spacetime
interpretation. The proof is due to Connes \cite{connes}. First using
$p^2=p$, we find $p[F, p]+ [F, p]p=[F, p]$ or
\begin{equation}
p[F, p] = [F, p]({\bf 1}-p), \quad ({\bf 1}-p)[F, p] = [F, p]p.
\label{twocomm}
\end{equation}
Also, using (\ref{twocomm}),
\begin{equation}
-p[F, p]^2= -p[F, p]^2 p = pP - (pFp)^2.
\end{equation}
Thus
\begin{eqnarray} 
(-1)^n Tr \gamma Q {\cal F}(\frac{1}{2}\Gamma(p)d\Gamma(p))^n &=&
(-1)^n Tr \gamma p P [F, p]^{2n} \\ 
&=& Tr \gamma P[p-(pFp)^2]^n
\end{eqnarray} 
\begin{equation} 
= Tr P\left[\frac{1+\gamma}{2}p - F_+^{\dagger}F_+ \right]^n - Tr P
\left[\frac{1-\gamma}{2}p - F_+ F_+^{\dagger} \right]^n.
\end{equation} 
The non-zero eigenvalues of $F_+^{\dagger} F_+$ and $F_+
F_+^{\dagger}$ are equal and of same multiplicity, as elementary
arguments show. So this last expression is the index of $F_+$ on
$[(1+\gamma)/2] Q A^{2^{N+1}}$($\equiv$ difference in the number of
zero modes of $F_+^{\dagger}F_+$ in $[(1+\gamma)/2] Q A^{2^{N+1}}$ and
$F_+ F_+^{\dagger}$ in $[(1-\gamma)/2] Q A^{2^{N+1}}$). That is just
$\pm N$and is also the difference in dimensions of $[(1 \pm \gamma)/2]
Q A^{2^{N+1}}$:
\begin{equation}
(-1)^n Tr \gamma Q {\cal F}(\frac{1}{2}\Gamma(p)d\Gamma(p))^n = \pm N.
\end{equation}

It remains to show (c), that is that $Tr \gamma Q {\cal
F}(\omega)^n$ is independent of $\hat{\omega}_1$.

Set
\begin{eqnarray} 
c_t &=& \frac{1}{2} \Gamma (d \Gamma) + t p \hat{\omega}_1 p, \\
{\cal F}(c_t) &=& dc_t + c_t^2.
\end{eqnarray}
Then ${\cal F}(c_1) = {\cal F}(\omega)$. Also ${\cal F}(c_t)$ fulfills
the Bianchi identity
\begin{equation}
d{\cal F}(c_t) + [c_t, {\cal F}(c_t)] = [F+c_t, {\cal F}(c_t)] = 0.
\end{equation}

Now
\begin{eqnarray}
\frac{d}{dt} Tr Q \gamma {\cal F}(c_t)^n &=& n Tr Q \gamma
\frac{d{\cal F}(c_t)}{dt} {\cal F}(c_t)^{n-1} \\
&=& n Tr Q \gamma \{F+c_t, p \hat{\omega}_1 p\} {\cal F}(c_t)^{n-1} \\
&=& n Tr Q \gamma \{F+c_t, p \hat{\omega}_1 p{\cal F}(c_t)^{n-1}\} \\
&=& -n Tr [F+c_t, Q \gamma p \hat{\omega}_1 p {\cal F}(c_t)^{n-1}] = 0.
\end{eqnarray} 
So
\begin{equation}
Tr Q \gamma {\cal F}(c_t)^n = Tr Q \gamma {\cal F}(c_t)^n|_{t=0} = Tr
Q \gamma [F, p]^{2n}
\end{equation}
as required.

\section{Final Remarks}
In this paper we have proposed a formulation of fuzzy physics using
cyclic cohomology. It relies especially on the theory of chiral
fermions of $S_F^2$ (with no fermion doubling) as elaborated in
\cite{ongoing}. Its distinct characteristic is the ease with which it
reproduces continuum topological features like instantons,
$\theta$-terms and axial anomaly. We remark in this context that we
did not explicitly write the $N \neq 0$ versions of all the actions in
Section 5. But that is easily done by changing the projector $P$ to
$Q$ and reinterpreting symbols like $\nabla$ and ${\cal F}(\omega)$.

This paper offers persuasive evidence that a combination of fuzzy
manifolds and cyclic cohomology can become a potent approach to
discretization of continuum physics.

There is overlap of this work with previous research \cite{grklpr1} on
fuzzy physics, especially as regards the treatment of chiral
anomalies. We conclude the paper with a brief comparison of the two
approaches. Peter Pre\v{s}najder had an essential role in its
composition. 

In the formalism of \cite{grklpr1}, a central role is played by a
spin-1/2 variable $z = (z_1, z_2), z \neq 0$. The spatial coordinates
$n_i$ of $S^2$ are identified with $(z^{\dagger} \sigma_i
z)/\sqrt{z^\dagger z}$ while the left- and right- chiral components of
the Dirac spinor on $S^2$ in the presence of a monopole field with
monopole number $2k = \pm N$ $(N\geq 0)$ are
\begin{eqnarray}	
\psi^{(+)}(z, z^*) &=& \sum_{|m|-|n|=2k-1} a^{(+)}_{mn} z^{*m} z^n, \\
\psi^{(-)}(z, z^*) &=& \sum_{|m|-|n|=2k+1} a^{(-)}_{mn} z^{*m} z^n.
\end{eqnarray}
A multi-index notation is being used with $m=(m_1, m_2)$, $z^{*m} =
z_1^{*m_1} z_2^{*m_2}$, $|m| = |m_1|+|m_2|$ etc.

The spinor $z$, after the normalization $z^{\dagger} z=1$, describes
the three-sphere $S^3$ as a Hopf fibration over $S^2$. Thus
$\psi^{(\pm)}$ are being represented here as functions of the twisted
principal bundle ${\mathbb C}^2 - \{0\}$ over $S^2$ (with structure
group $U(1) \times {\mathbb R}^1$, ${\mathbb R}^1$ being
dilatations). This is an important point. 

The description of the fuzzy sphere $S_F^2$ is achieved in
\cite{grklpr1} by replacing $z_\alpha$ and $z^*_\alpha$ by
annihilation and creation operators $\chi_\alpha$ and $\chi^*_\alpha$
with the only elementary non-vanishing commutator $[\chi_\alpha,
\chi^*_\beta] = \delta_{\alpha \beta}$. The fuzzy versions of the
chiral components $\psi^{(\pm)}$ of the Dirac field are
\begin{eqnarray}
f &=& \sum_{\stackrel{|m|-|n|=2k-1,}{|m|+|n| = 2j_0 -1}} a^{(+)}_{mn}
\chi^{*m} \chi^n, \label{fspinor1}\\ 
g &=& \sum_{\stackrel{|m|-|n|=2k+1,}{|m|+|n| = 2j_0 -1}} a^{(-)}_{mn}
\chi^{*m} \chi^n. \label{fspinor2}
\end{eqnarray}
where a restriction on $|m|+|n|$ has been introduced. Since $\chi$ and
$\chi^*$ transform as spinors, it eliminates all angular momenta $>
j_0 -1/2$. Further, the value of $|n|$ and the power of annihilators
in $f$ (or $g$) is $j_0 -k$ (or $j_0 - k -1$). [It is assumed the $j_0
\geq |k|+1$.] The domain of $f$ (or $g$) is accordingly restricted in
\cite{grklpr1} to the vectors $(\chi^*)^{j_0 - k}|0 \rangle$ (or
$(\chi^*)^{j_0 - k-1}|0 \rangle$) of angular momentum $(j_0 -k)/2$ (or
$(j_0 -k -1)/2$). [$|0\rangle$ is the vacuum annihilated by
$\chi_\alpha$.] With this restriction, $f$ and $g$ can be interpreted
as finite-dimensional matrices.

Note that the domains of $f$ and $g$ differ in their spinorial
character: if one has integral angular momenta, the other has half-odd
integral angular momenta. This suggests the introduction of
supersymmetry \cite{grklpr1}.

As angular momentum is being cut-off at the {\it same} value $j_0 -
1/2$ in both $f$ and $g$, each angular momentum in the fuzzy spinor
occurs with both chiralities. This technique of cut-off thus projects
out the analogue of the unwanted top mode in our work.

The relation between $j_0$ here and our $l$ can be found by comparing
the top total angular momenta:
\begin{eqnarray}
j_0 &=& 2l+k+1=2l \pm  N/2 +1 \quad {\rm for} \; f, \\
    &=& 2l+k=2l \pm N/2 \quad {\rm for} \; g. 
\end{eqnarray}	
Here the choice among the $\pm$ signs is determined by the sign of
$k$. Note how the value of $l$ increases by half a unit as we go from
$f$ to $g$.

Although until this point there is a good correspondence between the
two approaches, there is an important aspect besides supersymmetry
which differentiates the two. The chiral fields $\psi^{(\pm)}$ are
functions on the bundle ${\mathbb C}^2 -\{0\}$ for any instanton
number $2k$. They are like the wave functions for Dirac monopoles
\cite{BMSSold}. In particular there are no separate operators for
orbital angular momentum and spin in this formalism.

In contrast, in our approach, there are separate operators $L_i^{L,R}$
and ${\cal L}_i$ for characterizing orbital angular
momentum. Furthermore, in the sectors with instanton numbers $\pm N$,
we have to introduce $N$ ``isospin'' operators $\vec{\tau}^{(i)}/2$
and combine them with orbital angular momentum and spin to find the
total angular momentum $\vec{J}$. This construction is the analogue of
the similar construction \cite{jacreb,hastho} for the 't
Hooft-Polyakov monopoles. Thus our approach has a close correspondence
to the description of the latter on the ``sphere at infinity''.

Now in the {\it continuum}, it is possible to map one description to
the other in a well-understood way \cite{bmmnsz}. A similar
possibility for fuzzy physics has not been investigated. 

Although the two approaches seem to differ in this manner, both have a
Dirac and chirality operator which mutually anti-commute. That is
enough to guarantee the presence of zero modes responsible for chiral
anomaly. Thus we can see from (\ref{fspinor1}, \ref{fspinor2}) that
for $2k=N >0$ the minimum angular momentum $(|m|-|n|)/2 =(N-1)/2$
occurs only in $f$ (while the rest occur in both $f$ and $g$). The
corresponding zero modes of the Dirac operator are therefore of
positive chirality and multiplicity $N$. If $2k=-N<0$, these zero
modes are seen to have negative chirality, but the same
multiplicity. All this is exactly as we found.

Further studies contrasting the two methods would be useful to expose
their relative merits for particular problems.

\noindent {\bf Acknowledgments} T. R. Govindarajan, Badis Ydri, Xavier
Martin, Denjoe O'Connor and Peter Pre\v{s}najder offered us many good
suggestions during this work while Apoorva Patel told us about
\cite{luscher}. We thank them for their help. The idea of introducing
$g$ as in Section 5.4 arose from discussions with Denjoe. The work of
APB was supported in part by the DOE under contract number
DE-FG02-85ER40231.

\bibliographystyle{unsrt}

\end{document}